\documentclass[aps,prb,twocolumn,amsmath,amssymb,floatfix,superscriptaddress]{revtex4-2}
\usepackage{blindtext}
\usepackage{epsfig}
\usepackage{color}
\usepackage{amsmath}
\usepackage{amsfonts}
\usepackage{amssymb}
\usepackage{graphicx}%
\usepackage{url}
\usepackage[breaklinks]{hyperref}
\usepackage[english]{babel}
\usepackage{esvect}
\usepackage[normalem]{ulem}
\usepackage[inline]{enumitem}
\usepackage{xcolor}
\usepackage{xfrac}
\usepackage{dsfont}
\usepackage{bm}
\usepackage{array}
\newcolumntype{C}[1]{>{\centering\let\newline\\\arraybackslash\hspace{0pt}}m{#1}}
\usepackage{footnote}
\usepackage{isomath}
\DeclareMathAlphabet\mathbfcal{OMS}{cmsy}{b}{n}

\hypersetup{colorlinks,citecolor=blue,filecolor=blue,linkcolor=blue,urlcolor=blue}

\newcommand{\ket}[1]{\vert #1 \rangle}

\bibpunct{[}{]}{,}{n}{}{}

\begin{document}

\title{Majorana fermions and quantum information with fractional topology and disorder}
\author{Ephraim Bernhardt, Brian Chung Hang Cheung, Karyn Le Hur}
\affiliation{CPHT, CNRS, Institut Polytechnique de Paris, Route de Saclay, 91128 Palaiseau, France}

\begin{abstract}
The quest to identify and observe Majorana fermions in physics and condensed-matter systems remains an important challenge. Here, we introduce a qubit (spin-$1/2$) from the occurrence of two delocalized zero-energy Majorana fermions in a model of two spins-$1/2$ on the Bloch sphere within the fractional one-half topological state. We address specific protocols in time with circularly polarized light and the protection of this delocalized spin-$1/2$ state related to quantum information protocols. We also show how disorder can play a positive and important role allowing singlet-triplet transitions and resulting in an additional elongated region for the fractional phase, demonstrating the potential of this platform related to applications in topologically protected quantum information. We generalize our approach with
an array and Majorana fermions at the edges in a ring geometry.
\end{abstract}
\maketitle

\section{Introduction}

The study of Majorana fermions as effective excitations in condensed matter systems has spurred a lot of enthusiasm \cite{kitaev2001unpaired, alicea2012new, mi2022noise}. The interest in them is founded in potential applications for topological quantum computation \cite{nayak2008non}. While topologically protected Majorana fermions have been predicted in a p-wave superconductor \cite{read2000paired,kopnin1991mutual,volovik,kitaev2001unpaired}, Majorana zero modes can also occur in superconducting wires \cite{KLHthesis} as a result of the two-channel Kondo model and a magnetic impurity at an edge \cite{NozieresBlandin,EmeryKivelson}. Majorana zero modes can even occur in much smaller systems comprising only two sites \cite{leijnse2012parity}. Such a setup has recently been implemented experimentally with two quantum dots, showing agreement with the predicted physical properties of Majorana zero modes \cite{dvir2023realization}. These have been termed `poor man's Majorana bound states', as they do not enjoy topological protection.
This setup with two quantum dots is reminiscent of a system showing entangled fractional topological properties \cite{hutchinson2021quantum}. In particular, the Chern number defined from a quantum Hall effect in parameter space \cite{gritsev2012dynamical} for a spin in a radial magnetic field can fractionalize when several such spins interact. For two spins, there is a curious relation between the occurence of one-half partial topological numbers and free Majorana fermions \cite{hur2023one}. 
In this Article, we show the potential of this system with Majorana fermions as a non-local singlet-triplet qubit activated through circularly polarized light. 
We show that our approach can be generalized with an array and a pair of Majorana fermions at the edges, in a ring geometry with a weak link.

In Sec. \ref{topology}, we introduce the topological states on the sphere and recall the special aspect of the fractional $\frac{1}{2}$ state to realize a pair of zero-energy Majorana fermions delocalized between two spheres.
In Sec. \ref{Majorana}, we introduce the $\tau$-spin from these two delocalized Majorana fermions corresponding to singlet-triplet transitions. We address the positive role of disorder to realize a Pauli-X gate and generalize the fractional
state for a positive transverse $r_{xy}$ coupling between spins, introducing the disordered partial Chern marker. In Sec. \ref{quantuminformation}, as an application in quantum information, we navigate on the  
Bloch sphere throuch a circularly polarized field. We also generalize the analysis to a spin array showing that our approach for the $\tau$-spin can yet be applied in a ring geometry with one Majorana fermion at each edge. 
In Sec. \ref{protection}, in a Caldeira-Leggett model approach we show the protection of the $\tau$-spin towards dephasing associated to the dynamics of Majorana fermions. 

\section{Topological States on the Sphere}
\label{topology}

First, we introduce topological properties from one spin-1/2 in a radial magnetic field. 
The Hamiltonian reads
\begin{equation}
    \label{eq:Hrad}
    \mathcal{H}_\text{rad}= -d(\sin \theta \cos\phi \sigma^x+ \sin \theta \sin \phi \sigma^y + \cos \theta \sigma^z) - M \sigma^z.
\end{equation}
The parameter $M$ plays the role of a Semenoff mass \cite{semenoff1984condensed, hutchinson2021quantum}.
This model is a real space analogue of the Haldane model in momentum space \cite{haldane1988model} and has been used experimentally to probe the topological properties of the latter \cite{schroer2014measuring, roushan2014observation} with a dynamical protocol \cite{gritsev2012dynamical}. The Chern number can be defined in the parameter space with the polar and azimuthal angles $\theta$ and $\phi$ as
\begin{equation}
  \label{eq:Chern_general_definition}
  C = \frac{1}{2\pi} \int_\mathcal{S} dS_{\theta \phi} \mathcal{F}_{\theta \phi},
\end{equation}
where the Berry curvature $\mathcal{F}_{\theta \phi} = \partial_\theta A_\phi - \partial_\phi A_\theta$ is defined from the Berry connections $A_\mu = i \langle \psi| \partial_\mu |\psi\rangle$ and with $S_{\theta \phi} = d \theta d \phi$.
It can be shown that the Chern number for the model under consideration can be determined solely from the spin expectation values at the poles of the parameter space, i.e., \cite{henriet2017topology, hutchinson2021quantum}
\begin{equation}
    \label{eq:C_from_exp}
    C = \frac{1}{2} \left(\langle{\sigma^z(\theta = 0)}\rangle - \langle{\sigma^z(\theta = \pi)\rangle} \right).
\end{equation}
Next to provide a convenient and feasible platform to define topology in a quantum mechanical system, the model from Eq.~\eqref{eq:Hrad} is also interesting in its own rights for dynamical applications \cite{gritsev2012dynamical, henriet2017topology}, in particular with regards to energy conversion on the quantum scale \cite{bernhardt2023topologically}. Coupling several spins together gives rise to exotic fractional values of the effective topological number of each subsystem \cite{hutchinson2021quantum}. 

The model to realize fractional topology is identical to two spins with a radial magnetic field $\sum_{i=1}^2 {\cal H}_{\text{rad},i}$, as in Eq. (\ref{eq:Hrad}), in the presence of an Ising interaction $r_z\sigma_1^z \sigma_2^z$ \cite{hutchinson2021quantum,hur2023one}. The main mechanism to observe a fractional topological number on each {\it sub-system} is the formation of a Einstein-Podolsky-Rosen (EPR) pair at south pole, $|GS_S\rangle = \frac{1}{\sqrt{2}}(|\uparrow\downarrow\rangle + |\downarrow\uparrow\rangle)$ such that $\langle \sigma_i^z(\theta=\pi)\rangle=0$, whereas at north pole we have $|GS_N\rangle = |\uparrow\uparrow\rangle=|{\uparrow}_1\rangle \otimes |{\uparrow}_2\rangle$ and $\langle \sigma_i^z(\theta=0)\rangle=1$. Generalizing Eq.~(\ref{eq:C_from_exp}) for each spin this gives rise to a half topological number measurable when driving from north to south when activating the polar angle linearly in time $\theta=v t$ \cite{hutchinson2021quantum,hur2023one}. Important prerequisites to be fulfilled to observe one-half topological numbers on each sphere are the inversion symmetry between the two spins and $d-M<r_z<d+M$. This gives rise to an effective Hamiltonian close to the south pole \cite{hutchinson2021quantum}
\begin{equation}\label{eq:Heff_spin}
    \mathcal{H}_{\text{eff}}^{|{\uparrow \downarrow}\rangle,|{\downarrow \uparrow}\rangle}(\theta = \pi^-) = r_z \sigma_1^z \sigma_2^z - \frac{r_z d^2 \sin^2 \theta}{r_z^2 - (d-M)^2}\sigma_1^x \sigma_2^x,
\end{equation}
which fixes the EPR state as the ground state. 

To acquire an intuition about the occurrence of Majorana fermions in this system, it is useful to introduce the Jordan-Wigner transformation for two spins \cite{jordan1928paulische}. At the north pole, the ground state is described by one fermionic quasiparticle on each site through the identification $\sigma_i^z=2c^{\dagger}_i c_i -1= 1$, such that it satisfies $c^{\dagger}_i c_i |GS_N\rangle = |GS_N\rangle$. We implicitly assume that the spin eigenvalues are $\sigma_i^z=\pm 1$ such that $c^{\dagger}_i c_i=1,0$.

At the south pole, since we must satisfy $\langle GS_S|\sigma_i^z|GS_S\rangle=0$ for each sphere, this motivates a different transformation $\sigma_{1}^z=\frac{1}{i}(c^{\dagger}_1-c_1)$, $\sigma_{1}^x=(c^{\dagger}_1+c_1)$, $\sigma_{2}^z=\frac{1}{i}(c^{\dagger}_2-c_2)e^{i\pi c^{\dagger}_1 c_1}$ and $\sigma_{2}^x=(c^{\dagger}_2+c_2)e^{i\pi c^{\dagger}_1 c_1}$. We can then introduce the four Majorana fermions, $\eta_j = \frac{1}{\sqrt{2}}(c_j+c^{\dagger}_j)=\eta_j^{\dagger}$ and $\alpha_j = \frac{1}{\sqrt{2}i}(c^{\dagger}_j -c_j)=\alpha^{\dagger}_j$ with $2 i \eta_j \alpha_j = 1-2c^{\dagger}_j c_j$. This results in \cite{hur2023one,ReviewKLH2023}
\begin{equation}
  \label{Majoranaequation}
  H_{\text{eff}}^{|{\uparrow \downarrow}\rangle,|{\downarrow \uparrow}\rangle}(\theta = \pi^-) = -2ir_z \eta_1 \alpha_2 - \frac{ r_z d^2 \sin^2 \theta}{r_z^2 - (d-M)^2} 2i \alpha_1 \eta_2.
\end{equation}
Effectively, at $\theta=\pi$, we have two zero-energy (free) Majorana fermions $\alpha_1$ and $\eta_2$. They encode the entropy $2\times \frac{1}{2}\ln 2$ associated to the two classical ground states $|{\uparrow\downarrow\rangle}$ and $|{\downarrow\uparrow}\rangle$ before switching on the perturbation in $\sigma_{1}^x\sigma_{2}^x$ at $\theta=\pi-\epsilon=\pi^-$. The fractional topological number $C_j=\frac{1}{2}$ is stable and is measured indeed through a navigation from $\theta=0+\epsilon$ to $\pi-\epsilon$ on each sphere \cite{hutchinson2021quantum,hur2023one}. In the south pole region, we have $\langle GS_S| c^{\dagger}_i c_i |GS_S\rangle = \frac{1}{2}$. For comparison, when $r_z<d-M$ and $C_j=1$, each spin would rotate according to the radial magnetic field and therefore satisfy 
$\sigma_{i}^z=2c^{\dagger}_i c_i -1=-1$ at $\theta=\pi$ such that Majorana fermions would be paired on the same site, assuming that $d>M$. For $r_z>d+M$ and $C_j=0$, the ground states at the two poles are identical, i.e., $c^{\dagger}_i c_i |GS_S\rangle = c^{\dagger}_i c_i |GS_N\rangle$, such that the different Majorana fermions are then paired.

\begin{figure*}[t!]
  \centering
 \includegraphics[scale=0.35]{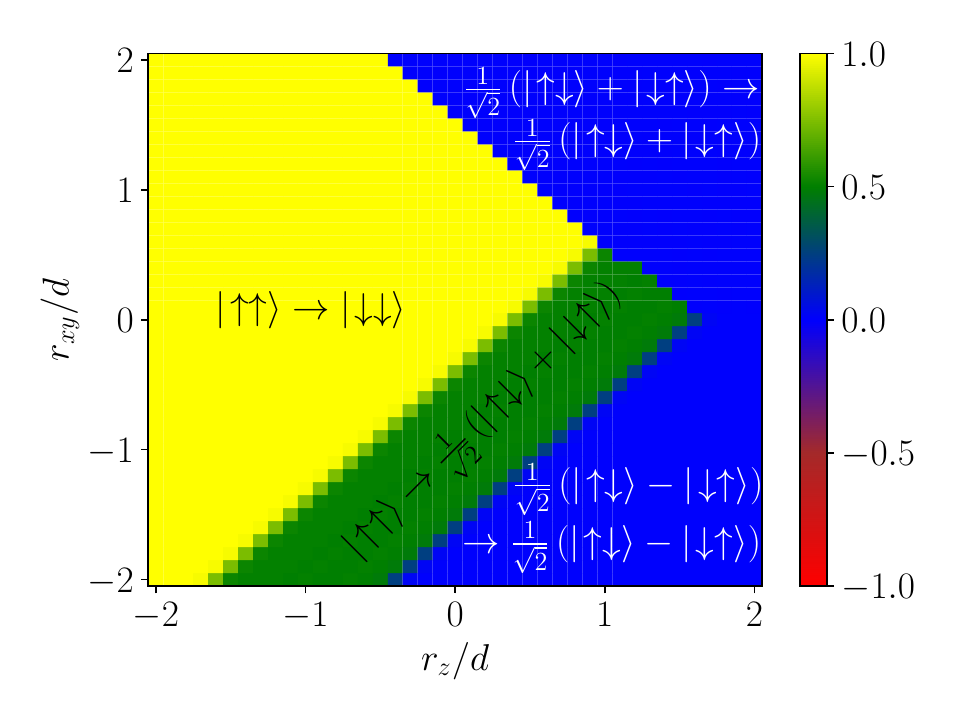}
 \includegraphics[scale=0.35]{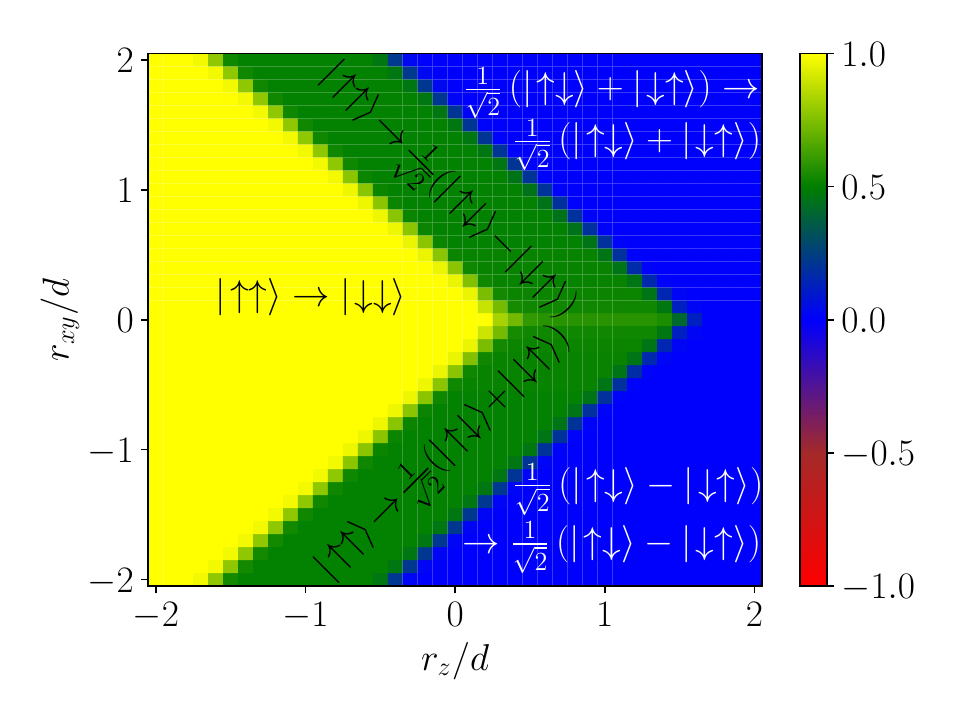}
 \vskip -0.2cm
  \caption{From \cite{bernhardtphd}. (Left) Clean phase diagram for $\delta M_1=\delta M_2=0$ showing $C_1$. (Right) Phase diagram with disordered $\delta M_1$ and $\delta M_2$ showing $\overline{C_1}$. Both phase diagrams were obtained using exact diagonalization (ED) for sphere $1$ when driving from north to south pole along the $\phi=0$ meridian with $d=1.0$, $M=0.6$ and $v=0.005$.
  The phase diagram with disorder was obtained by averaging the disordered partial Chern marker at each point in parameter space over $40$ different configurations of the mass disorder drawn from a normal distribution with $\overline{\delta M_i} = 0$ and variance $s_i = 0.1$. The transitions from north to south pole during a driving protocol $\theta = vt$ are written in the respective regions of the phase diagram.}
  \label{fig:phase_diagram}
\end{figure*}

\section{Majorana Fermions, $\tau$-Spin and Fractional Phase with Disorder}
\label{Majorana}

Here, we derive the $\tau$-spin from the two delocalized Majorana fermions and show that this is equivalent to a singlet-triplet qubit. We also show the role of disorder to realize a Pauli-X gate. Then, we present the generalized phase diagram in the presence of infinitesimal disorder introducing the disordered partial Chern marker. The fractional phase can then develop to the positive region of $r_{xy}$ couplings due to the fact that the $\tau^x$ operator now links singlet-triplet regions. 

\subsection{Majorana Fermions and $\tau$-spin}

The perturbative term in Eq.~\eqref{Majoranaequation} fixes a positive parity $2i\alpha_1\eta_2|GS_S\rangle = |GS_S\rangle$. However, acting with a transverse coupling term $2r_{xy}(\sigma_1^+\sigma_2^-+\sigma_1^-\sigma_2^+)$ can either support this positive parity or fix a negative parity. We can interpret these parities as the possible eigenvalues of a non-local two-level system, i.e., a qubit. In the basis of $\sigma$-spins, they correspond to the distinction of the singlet state $|{0,0}\rangle$ and the triplet state $|{1,0}\rangle$ in the added angular momentum basis.
The added angular momentum representation of these states is related to the tensor product representation by $|{1,1}\rangle=|{\uparrow \uparrow}\rangle$, $|{1,0}\rangle=\frac{1}{\sqrt{2}}\left(|{\uparrow\downarrow}\rangle+|{\downarrow\uparrow}\rangle\right)$, $|1,-1\rangle=|{\downarrow \downarrow}\rangle$ and $|{0,0}\rangle=\frac{1}{\sqrt{2}}\left(|{\uparrow\downarrow}\rangle-|{\downarrow\uparrow}\rangle\right)$. The triplet state $(1,0)$ corresponds to the parity $2i\alpha_1\eta_2|GS_S\rangle = |GS_S\rangle=+1$ and the singlet state
$(0,0)$ corresponds to $2i\alpha_1\eta_2|GS_S\rangle = |GS_S\rangle=-1$. When $\delta M_i=0$, a fractional topological state occurs when $\ket{1,0}$ is the ground state at south pole.

From the Majorana fermions $\alpha_1$ and $\eta_2$, we can define a non-local fermion $d=\frac{1}{\sqrt{2}}(\alpha_1+i\eta_2)$ and employing the Jordan-Wigner transformation in its usual form, we can introduce the non-local spin degrees of freedom 
$\tau^z=2d^{\dagger}d-1$, $\tau^x=d^{\dagger}+d$ and $\tau^y=-i(d^{\dagger}-d)$ with $ d^{\dagger} d =0,1$. We then find the description of the qubit in terms of the Majorana fermions, the non-local $d$-fermion and the $\sigma$-spins.
\begin{table}[h!]
\centering
\begin{tabular}{ c|c|c|c }
\hline
 $\tau$ & Majorana fermions  & $d$-operators & sphere operators\\
 \hline
 $\tau^z$ &  $2i\alpha_1\eta_2$&$2d^{\dagger}d-1$ & $\sigma_{1}^x\sigma_2^x$ \\
  $\tau^y$ & $-\sqrt{2}\eta_2$ & $-i(d^{\dagger}-d)$ & $\sigma_{2}^x\sigma_{1}^y$\\
 $\tau^x$ & $\sqrt{2}\alpha_1$ &$d^{\dagger}+d$ &  $\sigma_1^z$\\
 \hline
\end{tabular}
\end{table}\\
This defines the Hilbert space of the non-local $\tau$-qubit. In terms of the $\sigma$-spins, it is defined in the $\{ |{\uparrow \downarrow}\rangle,|{\downarrow \uparrow}\rangle \}$ subspace reached at the south pole. 
The triplet $|1,0\rangle$ and singlet $|0,0\rangle$ states correspond precisely to $\tau^z=\pm 1$ i.e. $2i\alpha_1\eta_2=\pm 1$. Here, we show the positive role of disorder for quantum information. A small mass imbalance (asymmetry) takes the form 
\begin{equation}
{\cal H}_{dis}=-\sum_i \delta M_i \sigma_i^z.
\end{equation}
If we use the identifications in Table 1, then we obtain $\sigma_1^z = \sqrt{2}\alpha_1$. Now, we also have the precise identity
\begin{eqnarray}
\sigma_2^z &=& \frac{1}{i}(c_2^{\dagger}-c_2)e^{i\pi c_1^{\dagger} c_1} = \sqrt{2}\alpha_2(1-2c^{\dagger}_1 c_1)  \nonumber \\
&=& \sqrt{2}\alpha_2(2i\eta_1\alpha_1).
\end{eqnarray}
In the presence of the antiferromagnetic interaction $r_z$, this is equivalent to navigate in the subspace $2i\eta_1\alpha_2=+1$. Therefore, this is equivalent to
\begin{equation}
\sigma_2^z = -\sqrt{2}\alpha_1.
\end{equation}
Through the implementation of the equality $2i\eta_1\alpha_2=+1$, the small disorder term takes the form
\begin{eqnarray}
{\cal H}_{dis} &=& - \delta M_1 \sigma_1^z - \delta M_2 \sigma_2^z = -\delta M_1 \sqrt{2}\alpha_1 +\delta M_2 \sqrt{2}\alpha_1 \nonumber \\
&=& (\delta M_2-\delta M_1)\sqrt{2}\alpha_1.
\end{eqnarray}
We also deduce
\begin{eqnarray}
{\cal H}_{dis} &=& (\delta M_2-\delta M_1)\sqrt{2}\alpha_1 = (\delta M_2-\delta M_1)(d+d^{\dagger})  \nonumber \\
&=&  (\delta M_2-\delta M_1)\tau^x. 
\end{eqnarray}
A small noise allows us to perform singlet-triplet transitions in the basis $\{|1,0\rangle;|0,0\rangle\}$: ${\cal H}_{s-t} =  (\delta M_2-\delta M_1)|1,0\rangle\langle 0,0| +h.c.$. Including a small $r_{xy}$ coupling, then this gives rise to
\begin{equation}
{\cal H}_{xy} = 2r_{xy}(\sigma_1^+ \sigma_2^- +\sigma_1^- \sigma_2^+) = r_{xy}(\sigma_1^x \sigma_2^x + \sigma_1^y \sigma_2^y) = 2r_{xy}\tau^z.
\end{equation}
The effect of such a coupling on the fractional partial topology has first been considered in \cite{hutchinson2021quantum}.
The $\tau$-spin is then described through the Hamiltonian
\begin{equation}
{\cal H}_{\tau}=(\delta M_2-\delta M_1)\tau^x + 2r_{xy}\tau^z.
\end{equation}
The positive role of infinitesimal disorder (noise) comes from the fact that it precisely allows us to navigate on the Bloch sphere and to realize singlet-triplet transitions:
\begin{equation}
{\cal H}_{\tau} = (\delta M_2-\delta M_1)\sqrt{2}\alpha_1 +2r_{xy}(2i\alpha_1\eta_2).
\end{equation}
We emphasize here that this form of Hamiltonian implicitly assumes that we are within the fractional phase of the model of two spheres, i.e. the $r_z$ coupling in this model equivalently fixes the parity state $2i\eta_1\alpha_2=+1$. We observe that the mapping between the $\tau$ spin and the Majorana fermions already reveals some protection, e.g. the $\tau^y$ operator has a non-local form in terms of the original spins operator which already ensures some protection. We will specifically address the question of the protection of the $\tau$ spin in Sec. \ref{protection}.

\subsection{Fractional Phase with $r_{xy}$ and Noise}

Here, we address the generalization of the phase diagram of the model of two spheres including small disorder in $M_1$ and $M_2$ taking into account that when $\delta M_i\neq 0$ this can now mediate triplet-singlet transitions.

At the south pole, for one disorder realization, this lifts the degeneracy between the states $\{|{\uparrow \downarrow}\rangle, |{\downarrow \uparrow}\rangle \}$. The ground state at south pole is changed into $|{\uparrow \downarrow}\rangle$ if $\delta M_1>\delta M_2>0$ and into $|{ \downarrow \uparrow}\rangle$ if $\delta M_2> \delta M_1>0$. The topology is measured using a dynamic protocol with $\theta = vt$ as in \cite{gritsev2012dynamical, schroer2014measuring, roushan2014observation}. With a small mass asymmetry, the fractional state will in practice still be observed when  $v \gg |(M_1 - M_2)|$. In order to remain close to the fractional topological ground state with disorder, we introduce a transverse coupling $r_{xy}$ between the spins \cite{hutchinson2021quantum,ReviewKLH2023}. 

In the limit $|\delta M_1-\delta M_2|\ll |r_{xy}|$, for $r_{xy}<0$ i.e. the triplet region at south pole, we verify
\begin{equation}
|GS_S\rangle \sim \left(|1,0\rangle+\frac{|\delta M_2-\delta M_1|}{-4r_{xy}}|0,0\rangle\right).
\end{equation}
This gives rise to
\begin{equation}
\langle GS_S | \sigma_j^z |GS_S\rangle = \mp \frac{|\delta M_2-\delta M_1|}{2r_{xy}}
\end{equation}
such that 
\begin{equation}
C_{1/2} = \frac{1}{2} \pm \frac{|\delta M_2-\delta M_1|}{4r_{xy}} = \frac{1}{2} \mp \frac{|\delta M_2-\delta M_1|}{4|r_{xy}|}.
\end{equation}
As long as the mass asymmetry remains much smaller than $4|r_{xy}|$ then the fractional $\frac{1}{2}$ state remains stable. Now, we study the positive region of $r_{xy}$
in the limit $0<|\delta M_1-\delta M_2|\ll r_{xy}$. For $r_{xy}>0$, then we obtain
\begin{equation}
|GS_S\rangle \sim \left(|0,0\rangle+\frac{|\delta M_2-\delta M_1|}{4r_{xy}}|1,0\rangle\right),
\end{equation}
such that
\begin{equation}
\langle GS_S| \sigma_{j}^z |GS_S\rangle = \pm \frac{|\delta M_2-\delta M_1|}{2r_{xy}}
\end{equation}
with $j=1,2$ respectively. This results in 
\begin{equation}
C_{1/2} = \frac{1}{2}\mp \frac{|\delta M_2-\delta M_1|}{4r_{xy}}.
\end{equation}
This equation then links the same fractional value for $C_i$ within the triplet ($r_{xy}<0$) and singlet ($r_{xy}>0$) regions. In Appendix \ref{disorder}
we verify that we reach a singlet ground state at south pole when driving in the adiabatic limit with the driving velocity being much smaller than the energy gap between the two lowest energy bands. 

Therefore, we observe that as long as $0<|\delta M_2-\delta M_1|\ll 4r_{xy}$, the singlet region of the phase diagram now also allows for a fractional number mimicking the $C_{1/2}\rightarrow 1/2$ case while maintaining the adiabatic driving. If we average on normal disorder realizations peaked around $\delta M_1=\delta M_2$ or $\delta M_i=0$ then the {\it disordered partial Chern marker} tends to $1/2$, as shown in Fig. \ref{fig:phase_diagram}. This result for the partial disordered Chern number $\overline{C}_i$ implies a finite variance $\sigma_i$ for each $\delta M_i$ such that $\overline{C}_i\rightarrow \frac{1}{2}$ if $\frac{\sigma_i}{4}\ll r_{xy}$ as shown in Fig. \ref{fig:phase_diagram} for $\frac{\sigma_i}{4}=0.025$. A small mass asymmetry produces a $\tau^x$ term in the Hamiltonian which unifies the $r_{xy}>0$ and $r_{xy}<0$ regions of the phase diagram. We emphasize here that in the clean limit, the energy gap between the two lowest energy bands goes to zero such that it is not possible to reach the singlet.

Below, we discuss applications for the realization of a non-local (delocalized) qubit formed with the Majorana fermions $\alpha_1$ and $\eta_2$. Applying a circularly polarized field, through perturbation theory, we show that it is indeed possible to realize the coupling $r_{xy}$
corresponding to a term $\tau^z$.

\section{Quantum Information}
\label{quantuminformation}

In this Section, we show that applying a circularly polarized electromagnetic wave we can navigate on the whole Bloch sphere linked to the $\tau$ spin. We also show that
the physics of the $\tau$ spin is equivalently obtained in an array generalizing the model in Eq. (\ref{eq:Hrad}).

\subsection{Manipulation with Circularly Polarized Light}

To realize e.g. a $\tau^y$ operation on the non-local $\tau$-qubit at the south pole ($\theta = \pi$), we implement a time-dependent protocol on the $\sigma$-spins by the Hamiltonian
\begin{equation}
    \label{eq:circ_light}
    V =A_1e^{-i \omega_1 t}\sigma_{1}^{+}+A_2e^{-i \omega_2 t}\sigma_{2}^{+}+\text{h.c.}.
\end{equation}
This corresponds to coupling to circularly polarized light if we render the two-level systems as dipoles \cite{klein2021interacting,le2022global} and has similarities to the momentum-space interaction Hamiltonian of a Haldane model with circularly polarized light \cite{tran2017probing, le2022global}. The drive from the north to the south pole here has the function of preparing a basis state of the $\tau$-qubit at the south pole.
The system is then held at the south pole and the time-dependent driving $V$ in Eq.~\eqref{eq:circ_light} is turned on. The effect of such a driving can most easily be understood in the rotating frame. We therefore apply a transformation $U=e^{i\frac{\omega_1 t}{2}\sigma_{1}^z} e^{i\frac{\omega_2 t}{2}{\sigma_{2}^z}}$. The spin Hamiltonian at $\theta = \pi$ commutes with this transformation, while the time-dependent driving transforms as
\begin{equation}
\label{eq:circ_light_rotating}
    \tilde{V} = A_1 \sigma_{1}^{+}+A_2\sigma_{2}^{+}+\text{h.c.} - \frac{\omega_1}{2}\sigma_1^z- \frac{\omega_2}{2}\sigma_2^z.
\end{equation}
In the $C_i = 1/2$ phase, we can take the driving term into account perturbatively. 

\begin{figure*}[ht]
    \centering
 \includegraphics[scale=0.37]{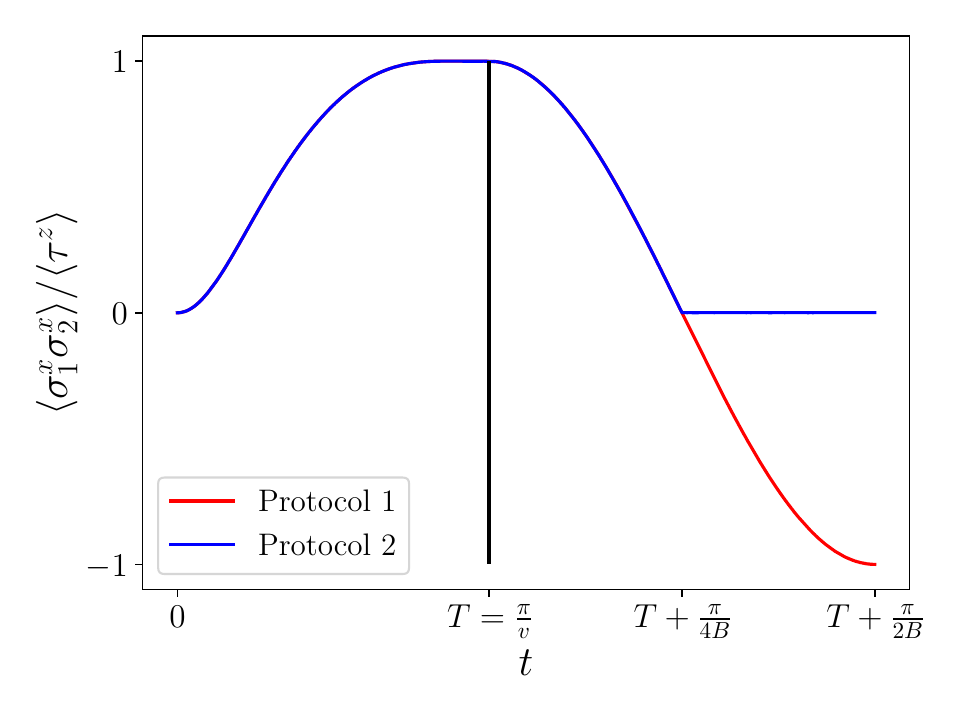}
 \includegraphics[scale=0.37]{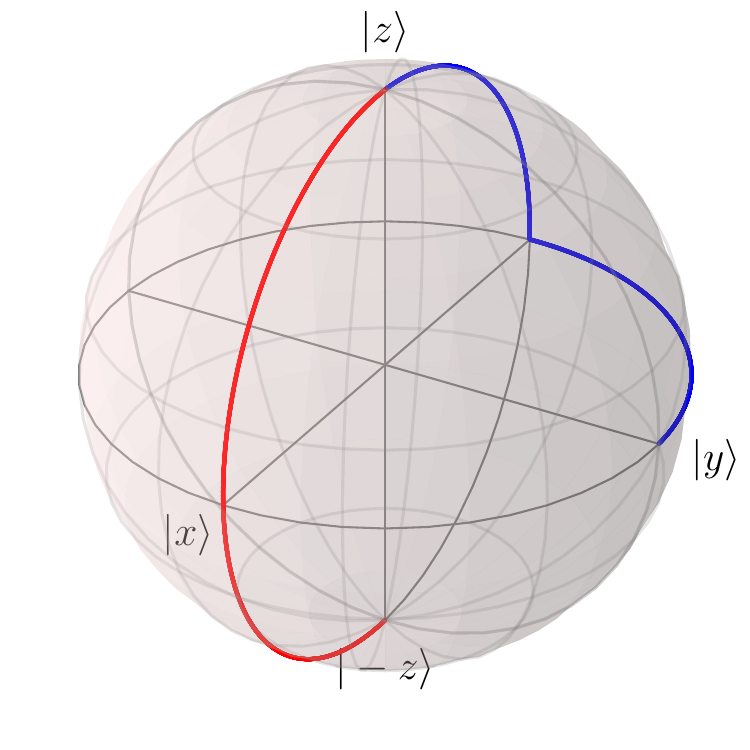}
 \vskip -0.2cm
  \caption{Here we implemented a driving from the north to the south pole with $d=1.0$, $M=0.9$, $r_z=1.0$ and $v=10^{-3}$.
  (Left) Parity $\langle \sigma_1^x \sigma_2^x\rangle=\langle \tau^z\rangle$ in time.  After reaching the south pole at $T=\pi/v$ on each sphere associated to the $\sigma$ spins, we activate the periodic driving protocol from Eq.~\eqref{eq:circ_light_rotating}. This allows us to access the Bloch sphere of the non-local $\tau$-qubit as shown by two exemplary paths. (Right) Path on the Bloch sphere associated to the $\tau$-spin starting from the time $T = \pi/v$ corresponding to the triplet state i.e. $\tau^z=+1$ (north pole).
  This allows us to access the Bloch sphere of the non-local $\tau$-qubit as shown by two exemplary paths. For the subsequent time-dependent protocol, we use $\omega_1 = \omega_2 = 0.01$ and for Protocol 1, represented by the red path, we set $A_1/d = -0.02i$ and $A_2/d = 0.02$, allowing to eventually reach the singlet state with $\langle{\tau^z}\rangle = -1$. For Protocol 2, represented by the blue path, we first set $A_1/d = 0.02i$ and $A_2/d = 0.02$ for the first half of the protocol, while for the second half we set $A_1/d = 0.02$ and $A_2/d = 0.02$. This allows us to reach the state with $\langle {\tau^y}\rangle = 1$.}
  \label{fig:gates}
\end{figure*}

In the rotating frame, the full Hamiltonian at the south pole reads
\begin{equation}
    \mathcal{H} = \mathcal{H}_0 + \tilde{V},
\end{equation}
where $\mathcal{H}_0=-(d-M)(\sigma_1^z+\sigma_2^z)+r_z\sigma_1^z\sigma_2^z$. For simplicity, we begin with $r_{xy}=0$ here.
We denote the degenerate ground state subspace $\{|{\uparrow\downarrow}\rangle, \langle{\downarrow\uparrow}\rangle \}$ by $a$, the subspace $\{|{\uparrow\uparrow}\rangle\}$ by $b$ and the subspace $\{|{\downarrow\downarrow}\rangle\}$ by $c$.
Degenerate perturbation theory in the ground state subspace then gives the effective Hamiltonian
\begin{equation}\label{eq:eff_ham}
    \mathcal{H}_\text{eff}=\mathcal{H}_a+\tilde{V}_a+\sum_{l}\tilde{V}_{al}\frac{1}{E_a-E_l}\tilde{V}_{la},
\end{equation}
where $\mathcal{H}_a$ is the unperturbed Hamiltonian in the $a$ subspace, $\tilde{V}_a$ is the perturbation Hamiltonian in the $a$  subspace and $\tilde{V}_{ab}=P_a\tilde{V}P_b$, where $P_a$ and $P_b$ are the projectors onto the $a$ and $b$ subspaces respectively.
We then find
\begin{eqnarray*}
    \mathcal{H}_a &=& -r_z(|{\uparrow\downarrow}\rangle \langle {\uparrow\downarrow}|+|{\downarrow\uparrow}\rangle \langle {\downarrow\uparrow}|),\\
    \tilde{V}_a &=&\frac{\omega_2-\omega_1}{2}(|{\uparrow\downarrow}\rangle\langle {\uparrow\downarrow}|-|{\downarrow\uparrow}\rangle\langle {\downarrow\uparrow}|),\\
    \tilde{V}_{ab}&=& A_1^{*}|{\downarrow\uparrow}\rangle\langle {\uparrow\uparrow}|+A_2^{*}|{\uparrow\downarrow}\rangle\langle {\uparrow\uparrow}|,\\
   \tilde{V}_{ac}&=& A_1|{\uparrow\downarrow}\rangle\langle {\downarrow\downarrow}|+A_2|{\downarrow\uparrow}\rangle\langle {\downarrow\downarrow}|.
\end{eqnarray*}
The second order perturbation term in matrix form with the basis $a$ reads
\onecolumngrid
\begin{equation}
    \begin{split}
        \sum_{l}\tilde{V}_{al}\frac{1}{E_a-E_l}\tilde{V}_{la}&=\begin{pmatrix}
                    \frac{|{A_2}|^2}{E_a-E_b}+\frac{|A_1|^2}{E_a-E_c} & (\frac{1}{E_a-E_b}+\frac{1}{E_a-E_c})A_2^{*}A_1 \\
                    (\frac{1}{E_a-E_b}+\frac{1}{E_a-E_c})A_2A_1^{*} & \frac{|A_1|^2}{E_a-E_b}+\frac{|A_2|^2}{E_a-E_c}
                    \end{pmatrix}\\
                  &=\begin{pmatrix}
                    \frac{-r_z(|{A_1}|^2+|A_2|^2)+(d-M)(|A_1|^2-|{A_2}|^2)}{2r_z^2-2(d-M)^2} & \frac{-r_z}{r_z^2-(d-M)^2}A_2^{*}A_1 \\
                    \frac{-r_z}{r_z^2-(d-M)^2}A_2A_1^{*} & \frac{-r_z(|A_1|^2+|A_2|^2)-(d-M)(|A_1|^2-|A_2|^2)}{2r_z^2-2(d-M)^2}
                    \end{pmatrix}.\\
    \end{split}
\end{equation}
\twocolumngrid
Noting that $E_a=-r_z$, $E_b=2(d-M)+r_z$ and $E_c=-2(d-M)+r_z$ and using the Pauli matrix representation of the $\tau$-spin, we find
\begin{equation}
    \tilde{V}_\text{eff}=B_0\mathbb{I}_2+\mathbf{B}\cdot\bm{\tau},
\end{equation}
where $B_0=\frac{-r_z(|A_1|^2+|A_2|^2)}{2r^2-2(d-M)^2}-r_z$ and $\mathbf{B}=(B_x, B_y, B_z)$,
with
\begin{subequations}
  \label{exact_B}
\begin{eqnarray}
        B_x &=&\frac{(d-M)(|A_1|^2-|A_2|^2)}{2r_z^2-2(d-M)^2}+\frac{\omega_2-\omega_1}{2},\\
        B_y &=&\frac{-r_z}{r_z^2-(d-M)^2}\hbox{Im}{A_2^{*}A_1},\\
        B_z &=&\frac{-r_z}{r_z^2-(d-M)^2}\hbox{Re}{A_2^{*}A_1}.
    \end{eqnarray}
\end{subequations}

Tuning the amplitudes $A_i$ and the driving frequencies $\omega_i$, we see that in the rotating frame, we can reach all states on the Bloch sphere of the $\tau$-qubit. We implicitly assume $A_i\ll d$ to ensure the robustness of the fractional topological phase. We demonstrate two such manipulations thus playing the role of Pauli gates in Fig.~\ref{fig:gates}. Note that there is a precise correspondence between observables in the rotating and laboratory frames. For $\omega_1=\omega_2$, observables are identical in both frames which is then practical to navigate from one space to another within this scheme. Since the fractional phase implies $r_z\gg (d-M)$ in principle for $\omega_1=\omega_2$ we have $|B_z|\gg |B_x|$ such that this will help stabilizing the fractional phase with $B_z$ playing the role of $r_{xy}$.

\subsection{$\tau$-spin and Majorana fermions with a Ring}

Here, we show that the approach for the $\tau$ spin can also be adapted for a wire or an array, linking with a pair of Majorana fermions at the edges in the quantum Ising model or p-wave superconducting wire \cite{kitaev2001unpaired, alicea2012new, mi2022noise}.
Implementing this model in a ring, this is a step towards the elaboration of protected quantum platforms with our approach. We can generalize the Jordan-Wigner transformation as 
\begin{eqnarray}
\sigma_j^z &=& \frac{1}{i}(c^{\dagger}_j-c_j)e^{i\pi \sum_{k<j} c^{\dagger}_j c_j} \\ \nonumber
\sigma_j^x &=& (c_j+c_j^{\dagger})e^{i\pi\sum_{k<j} c^{\dagger}_j c_j}.
\end{eqnarray}
This results in
\begin{eqnarray}
    \mathcal{H}_{\text{eff}} &=& \sum_{j=1}^N r_z \sigma_j^z \sigma_{j+1}^z - \frac{r_z d^2 \sin^2 \theta}{r_z^2 - (d-M)^2}\sigma_j^x \sigma_{j+1}^x \\ \nonumber
 \mathcal{H}_{\text{eff}} &=& \sum_{j=1}^N -r_z 2 i \eta_j\alpha_{j+1} - \frac{r_z d^2 \sin^2 \theta}{r_z^2 - (d-M)^2} 2 i \alpha_j \eta_{j+1}.
\end{eqnarray}
We introduce the same Majorana fermions on each site $j$ as the ones introduced earlier $\{\alpha_j,\eta_j\}$. 
Then, we observe that since the transverse field goes to zero at $\theta=\pi$ in the two-spheres' model this stabilizes two edge Majorana fermions $\alpha_1=\frac{1}{\sqrt{2}i}(c^{\dagger}_1-c_1)=\alpha_1^{\dagger}$ and $\eta_N=\frac{1}{\sqrt{2}}(c_N+c_N^{\dagger})=\eta_N^{\dagger}$ encoding the double degeneracy of the classical Ising ground states. 
We assume here an even number of sites such that 
\begin{equation}
\sum_{j=1}^N (d-M)\sigma_j^z=0.
\end{equation}
For an even number of spins, the system shows the same topological number $C_j=\frac{1}{2}$ related to the formation of Greenberger-Horne-Zeilinger entangled states \cite{hutchinson2021quantum} as a result of the perturbations at $\theta=\pi^-$, i.e. when switching on the transverse field effects. On a chain with a finite number of sites, this allows to realize quantum superpositions with the two classical Ising ground states at angle $\theta=\pi^-$. (It is interesting to note that the term in $\sigma_j^x \sigma_{j+1}^x$ does not alter the 
presence of the two edge modes as long as $\sin\theta\ll 1$.) 
The edge Majorana fermion at $j=1$ can yet be related to a local spin observable 
\begin{equation}
\sigma_1^z = \sqrt{2}\alpha_1.
\end{equation}

To produce a delocalized qubit i.e. the $\tau$-spin, we can place the system in a ring geometry with an adjustable junction. In this way, in a ring geometry $\sigma_1^{\mu} = \sigma_{N+1}^{\mu}$ imply that 
\begin{equation}
e^{i\pi\sum_{1\leq j\leq N} c^{\dagger}_j c_j }=  \prod_{1\leq j\leq N} e^{i \pi c^{\dagger}_j c_j}=1. 
\end{equation}
In our case, $c^{\dagger}_j c_j$ is related
to the operator $\sigma^y_j=2c^{\dagger}_j c_j - 1=2i\alpha_j\eta_j$. In a ring geometry, the string may allow the phase to adjust in a way to account for the periodic boundary conditions. 
This is also equivalent to 
\begin{eqnarray}
e^{i\pi \sum_{1\leq j < N} c^{\dagger}_j c_j } &=&  \prod_{1\leq j< N} e^{i \pi c^{\dagger}_j c_j} \nonumber \\
&=& \left(\prod_{1\leq j\leq N} e^{i \pi c^{\dagger}_j c_j}\right)\times e^{i\pi c^{\dagger}_N c_N} \nonumber \\
&=& (1-2c^{\dagger}_N c_N).
\end{eqnarray}

Therefore, we have the identities
\begin{eqnarray}
\sigma_N^x \sigma_N^y &=& \sqrt{2}\eta_N e^{i\pi\sum_{1\leq j<N} c^{\dagger}_j c_j}(2c^{\dagger}_N c_N-1) \nonumber \\
&=& \sqrt{2}\eta_N (1-2c^{\dagger}_N c_N)(2c^{\dagger}_N c_N-1) \nonumber \\
&=& -\sqrt{2}\eta_N.
\end{eqnarray}
For fermions $(c^{\dagger}_j c_j)^2=c^{\dagger}_j c_j$.  Similarly,
\begin{eqnarray}
\sigma_{1}^z \sigma_N^z &=& \sigma_{N+1}^z \sigma_N^z = \sqrt{2}\alpha_1(1-2c^{\dagger}_N c_N)\sqrt{2}\alpha_N \nonumber \\
&=& 2\alpha_1(2i\eta_N\alpha_N)\alpha_N=2i\alpha_1\eta_N.
\end{eqnarray}
This term can be implemented through a small time-dependent Ising coupling $r'_z(t) \sigma_{N+1}^z \sigma_N^z$ between sites $1$ and $N$ in a (small) ring geometry. When $r_z=r'_z>0$ this corresponds to a chain with periodic boundary conditions and therefore 
there is no free Majorana fermions. The operator $2i\eta_N\alpha_1=2i\eta_N\alpha_{N+1}$ would remain equal to $+1$. If now, we have a small amplitude $|r'_z(t)|\ll r_z$ this may in principle allow to address both signs of the interaction in time while preserving the ground state in the bulk. This could then allow to flip the state of the parity operator $2i\eta_N\alpha_1=\pm 1$ in time without modifying the Majorana ground state in the bulk related to the Ising chain.

Then, we have the identifications
\begin{eqnarray}
\tau^x &=& \sqrt{2}\alpha_1 = \sigma_1^z \\ \nonumber
\tau^y &=& - \sqrt{2}\eta_N = -\sigma_N^x \sigma_N^y \\ \nonumber
\tau^z &=& 2i\alpha_1\eta_N = \sigma_1^z \sigma_N^z.
\end{eqnarray}
A small noise on the mass term on the first sphere $\delta M_1$ yet allows to activate the $\tau^x$ gate. The operators $\sigma_N^x$ and $\sigma_N^y$ are not relevant operators, but developing the partition function this may allow for the induction of a term 
in $\sigma_N^x \sigma_N^y$ through the presence of local fields at site $N$ along $x$ and $y$ axis. Coupling to classical light or a cavity, in general, may also allow to activate the $\tau$-spin.

\section{Protection}
\label{protection}

In this Section, we address the protection of the $\tau$-spin towards dephasing (decoherence) related to the dynamics of Majorana fermions.

\subsection{Caldeira-Leggett Model and $\tau$ spin}

A common way to model decoherence (dephasing) is  through the class of Caldeira-Leggett models \cite{leggett1987dynamics}, which also gives rise to the so-called spin-boson model with the Hamiltonian \cite{weiss2012quantum}
\begin{equation}
\label{spinbathcoupling}
    {\cal H} =  {\cal H}_\text{spin} + \frac{\sigma^z}{2} \sum_k g_k (b_k^\dag + b_k) + \sum_k \omega_k b_k^\dag b_k.
\end{equation}
For simplicity, the Planck constant $\hbar=h/(2\pi)$ is fixed to unity.
The environment (or bath) is modelled by an infinite number of quantum harmonic oscillators, characterized by its spectral function \cite{weiss2012quantum}
$J(\omega) = \pi \sum_k g_k^2 \delta(\omega-\omega_k)$.
It is well known that coupling to an environment leads to decoherence of the spin, manifest in the vanishing of the off-diagonal elements of the spin density matrix proportional to $\langle{\sigma^\pm(t)}\rangle$, which can be understood through a unitary transformation on the Hamiltonian (\ref{spinbathcoupling}) \cite{lehur2018driven}. It is interesting to observe though that the topological number as formulated from the magnetizations at the poles in Eq. (\ref{eq:C_from_exp}) is in general robust to decoherence for relatively large couplings with the bath  \cite{henriet2017topology}. 

When studying the decoherence of the $\tau$-qubit this requires to begin with the original two-spins model.
Therefore, we set $\mathcal{H}_\text{spin} = \sum_{i=1}^2\mathcal{H}_{\text{rad},i}(\theta = \pi) + r_z \sigma_1^z \sigma_2^z + V$ and the interaction term becomes $\sum_i \frac{\sigma_i^z}{2} \sum_k g_{k,i}(b_k^\dag+b_k)$.
In the $C_i = 1/2$ phase, the ground state at the south pole is found in the degenerate $\{ |{\uparrow \downarrow}\rangle, |{\downarrow \uparrow}\rangle \}$ subspace. Within this sub-space, the term $r_z$ will fix the parity operator $2i\eta_1\alpha_2=+1$ independently of the dynamics of the variables forming the $\tau$-spin. Here, one may ask about the roles of possible additional couplings of each individual spin with the bath along $x$ or $y$ direction. It is interesting to comment that $\sigma_1^x=\sqrt{2}\eta_1$ and that this Majorana fermion is
bound to the Majorana fermion $\alpha_2$ through the parity state $2i\eta_1\alpha_2=+1$ as a result of the $r_z$ Ising coupling. Similarly, $\sigma_2^x=\sqrt{2}\eta_2(2i\eta_1\alpha_1)$ and when navigating on the Bloch sphere activating the parity of the $\tau$-spin as
$2i\alpha_1\eta_2=\pm 1$, then $\sigma_2^x$ is also attached to the same Majorana fermion $\eta_1$. For the same reason the operator $\sigma_i^y$ cannot couple easily to the bath. Therefore, the Caldeira-Leggett model in Eq. (\ref{spinbathcoupling}) seems justified in the present case.
The interaction term with the bath can then be rewritten as 
\begin{equation}
\frac{\tau^x}{2}\sum_k (g_{k,1}-g_{k,2})(b_k^\dagger +b_k),
\end{equation}
with $\tau^x=\sqrt{2}\alpha_1$. From our definitions in the Table, $\tau^z$ is related to the diagonal part of the spin density matrix, therefore dephasing refers to the behavior of the $\tau^x$ and $\tau^y$ components.
The protection towards dephasing of the $\tau$ spin can be seen from the fact that the $\tau^y$ operator, i.e. $\sigma_2^x\sigma_1^y$, would not interact easily with a local noise term. In addition, the $\tau^x$ operator is linearly coupled to the bath whereas dephasing would imply a coupling to an operator that would exponentially depend on the operators $b_k$ and $b^{\dagger}_k$ \cite{lehur2018driven}. 
The non-locality of the $\tau^y$ operator in terms of original spins and the fact that the bath now couples longitudinally along $x$ direction with the bath
then justify the protection of the $\tau$-spin from dephasing.

\subsection{Protection and Dynamics}

Here, we describe the protection towards dephasing with Heisenberg equations of motion where we will precisely address the role of Majorana fermions. We will also compare the results with those of the usual spin-boson model. 
Suppose we begin with the Hamiltonian including the bath effect when $B_y=0$
\begin{eqnarray}
{\cal H}_{\tau} &=& B_x\tau^x+(\delta M_2-\delta M_1)\tau^x+2r_{xy}\tau^z \nonumber \\
&+&\sum_k \lambda_k(b_k+b_k^{\dagger})\tau^x +\sum_k \omega_k b^{\dagger}_k b_k,
\end{eqnarray}
with a (small) transverse field $B_x\neq 0$. The term $B_z$ can be generally included in the $r_{xy}$ term. We introduce $\lambda_k$ such that $\lambda_k=\frac{g_{k,1}-g_{k,2}}{2}$
We can then combine the noise effect and the bath term through a transverse field
\begin{equation}
\Delta = B_x+(\delta M_2-\delta M_1)+\sum_k \lambda_k(b_k+b_k^{\dagger}).
\end{equation}
We can write down the Heisenberg equations of motion 
\begin{eqnarray}
\label{dynamics}
\sqrt{2}\dot{\alpha}_1 &=& \dot{\tau}^x = -4 r_{xy}\tau^y = 4\sqrt{2}r_{xy}\eta_2  \nonumber \\
-\sqrt{2}\dot{\eta}_2 &=& \dot{\tau}^y = 4r_{xy}\tau^x - 2\Delta\tau^z \nonumber \\
&=& 4r_{xy}\sqrt{2}\alpha_1 - 2\Delta(2 i\alpha_1\eta_2).
\end{eqnarray}

From general grounds, the parity operator $2i\alpha_1\eta_2$ is real. Therefore, suppose we multiply the left-hand side of the first line in Eqs. (\ref{dynamics}) by $i\eta_2$ and the second line by $i\alpha_1$ these two equations become equivalent to
\begin{eqnarray}
i\eta_2\dot{\alpha}_1 &=& i4 r_{xy}\eta_2^2 \\ \nonumber
i\alpha_1\dot{\eta}_2 &=& - 2\sqrt{2}\Delta\alpha_1^2\eta_2 - 4i r_{xy}\alpha_1^2. 
\end{eqnarray}
This leads to 
\begin{equation}
i\frac{d}{dt}(\alpha_1\eta_2) = 4i r_{xy}(\eta_2^2-\alpha_1^2) - 2\sqrt{2}\Delta\alpha_1^2 \eta_2.
\end{equation}
Majorana fermions satisfy that $\eta_2^2=\alpha_1^2=\frac{1}{2}$. Therefore, from the differentials, this equation becomes
\begin{equation}
\label{dynamicsMajorana}
i\frac{d}{dt}(\alpha_1\eta_2) = - \Delta\sqrt{2}\eta_2.
\end{equation}

(i) Eq. (\ref{dynamicsMajorana}) agrees with the Heisenberg equation of motion $\dot{\tau}^z = 2\Delta\tau^y$. In general, this equation leads to a unitary evolution in time. Suppose that $\alpha_1(t)$ and $\eta_2(t)$ would acquire a decoherence such that $\alpha_1(t)\rightarrow \alpha_1.e^{-t/\tau_1}$ and $\eta_2(t)\rightarrow \eta_2.e^{-t/\tau_2}$, then on the left-hand side we would have a function in $e^{-\frac{t}{\tau_1} - \frac{t}{\tau_2}}$ whereas on the right-hand side we have a function in $e^{-t/\tau_2}$. It is also interesting to comment that the $\sigma^z$ variable related to the usual spin-boson model satisfies a similar equation as for the $\tau^z$ spin
\begin{equation}
\label{sigmaz}
\dot{\sigma}^z = 2\Delta\sigma^y.
\end{equation}
The correspondence between the spin-boson models with the $\sigma$ spin (of one sphere) and with the $\tau$ spin can be understood as $\sigma^z\rightarrow \tau^x$ (corresponding to the variable coupling to the bath), $\sigma^y\rightarrow \tau^y$ and $\sigma^x\rightarrow -\tau^z$. In terms of the Majorana fermions $\alpha_1$ and $\eta_2$, Eq. (\ref{sigmaz}) would imply $\dot{\alpha}_1 = - 2\Delta\eta_2$ such that we verify that a solution with an exponential decoherence would now be possible. 
We also verify the occurrence of decoherence when evaluating $\ddot{\sigma}^z$ to second-order in $\lambda_k$. Therefore, all these results agree with the fact that the Majorana fermions $\alpha_1$ and $\eta_2$ can admit a coherent unitary dynamics when the bath is linearly coupled to the $\tau$-spin along $x$ direction. Below, we verify the coherent dynamics in the $x-y$ plane for the $\tau$-spin. For completness, we can also address the fact that $\Delta$ is yet an operator.  Plugging Eq. (\ref{dynamicsMajorana}) into Eqs. (\ref{dynamics}), we obtain
\begin{equation}
\label{equationeta2}
\ddot{\eta}_2 = -4r_{xy}\dot{\alpha}_1 - 4\Delta^2\eta_2 = \left(-16 r_{xy}^2 - 4\Delta^2\right)\eta_2.
\end{equation}

If the bath is prepared within the vacuum state, in the weak-coupling limit, when the noise is infinitesimal then $\Delta^2$ is approximately equal to $B_x^2$. More precisely, from the equations of motion we also have
{\begin{equation}
\label{dynamicsbosons}
\ddot{b}_k +\ddot{b}^{\dagger}_k = (i\omega_k)^2(b_k+b_k^{\dagger}) + 2\lambda_k \omega_k\tau^x.
\end{equation}
If we begin with $\tau^x=0$ and $\tau^y=1$, then this equation leads to a free evolution of the bosons $b_k(t)=b_k(0)e^{-i\omega_k t}$ and $b_k^{\dagger}(t) = b_k^{\dagger}(0) e^{i \omega_k t}$. In the weak-coupling limit between spin and bath,
\begin{eqnarray}
\Delta^2 &\approx& B_x^2 + 2B_x\left((\delta M_2-\delta M_1)+\sum_k \lambda_k \langle b_k(t) + b_k^{\dagger}(t) \rangle\right) \nonumber \\
 &\sim& B_x^2.
\end{eqnarray}
\twocolumngrid
In the situation of unitary evolution for the bosons, then we yet have $\langle b_k(t) + b_k^{\dagger}(t)\rangle=0$ from the vacuum state. 
Therefore, Eq. (\ref{equationeta2}) leads to
\begin{equation}
\eta_2(t) = \eta_2(0) \cos\left(4\sqrt{r_{xy}^2 + \frac{B_x^2}{4}} t \right),
\end{equation}
with $\eta_2(0) = - \frac{1}{\sqrt{2}}$. From the equation $\dot{\alpha}_1 = \frac{4}{\hbar}r_{xy}\eta_2$ then we also obtain
\begin{equation}
\alpha_1(t) = \eta_2(0) \frac{r_{xy}}{\sqrt{r_{xy}^2 + \frac{B_x^2}{4}}}\sin\left(4\sqrt{r_{xy}^2 + \frac{B_x^2}{4}}t \right).
\end{equation}
We implicitly assume that $r_{xy}\gg B_x$ such that we remain within the fractional phase. 
These equations also reveal that $\tau^y$ is not coupled to the bath. If we begin with $\langle \tau^x\rangle \neq 0$, then $\alpha_1(t)$ will yet show an oscillation in time that will then give an additional term of the same nature to $b_k(t)+b_k^{\dagger}(t)$. This will result in fluctuations of $\Delta^2(t)$ around $B_x^2$. However, this does not lead to dephasing. Therefore, as long as $\lambda_k(b_k(t)+b^{\dagger}_k(t))\ll B_x$, then the Majorana fermions will yet show an oscillatory evolution in time with frequency $\sim 4\sqrt{r_{xy}^2+\frac{B_x^2}{4}}$.

(ii) These facts can be generalized for multiple baths coupling to $\sigma_1^z$ and $\sigma_2^z$ and in fact do not depend on the precise form of $\Delta$ in the sense that similarly to $(\delta M_2-\delta M_1)$ we can introduce 
$\sum_k \lambda_k^2(b_{2k}+b_{2k}^{\dagger}) - \sum_k \lambda_k^1(b_{1k}+b_{1k}^{\dagger})$ as two independent baths.  

(iii) If we include a term $B_y \tau^y$, we obtain
\begin{equation}
\dot{\alpha}_1 = 4 r_{xy}\eta_2 + \sqrt{2}B_y(2i\alpha_1\eta_2),
\end{equation}
and the equation for $\dot{\eta}_2$ remains identical. If we prepare the system at time $t=0$ with either $\alpha_1(0)=\frac{1}{\sqrt{2}}$ and $\eta_2(0)=0$ or $\alpha_1(0)=0$ and $\eta_2(0)=\pm\frac{1}{\sqrt{2}}$, then from the differentials the conclusions remain identical as above. These preparations are in fact acceptable through the coupled differential equations because correspond to sine waves for one Majorana fermion and cosine wave for the other Majorana fermion. 

\section{Conclusion}

We have studied the stability of the recently found $C_i = 1/2$ phase in interacting spins with regards to disorder and proposed an application for the preparation and activation of a non-local qubit, in a similar spirit as recent experimental realizations \cite{dvir2023realization}. Through this platform, we found that a transverse interaction can stabilize the fractional topological phase against disorder effects. The latter even gives rise to an extension of the fractional topological phase thanks to the lifting of a band crossing. Applying a mapping to Majorana fermions \cite{hur2023one}, we proposed an application for storing quantum information and showed a way how to manipulate it using a time-dependent protocol. Our work opens perspectives for applications in quantum information and the synthetization of Majorana bound states.

We acknowledge interesting discussions with Landry Bretheau, Olesia Dmytruk and Clement Roux. This work was supported by the French ANR BOCA and the Deutsche Forschungsgemeinschaft (DFG), German Research Foundation under Project No. 277974659.

\begin{appendix}
\section{Triplet-Singlet Transition and Disorder}
\label{disorder}

\begin{figure*}[ht!]
  \centering
      \includegraphics[width=0.3\textwidth]{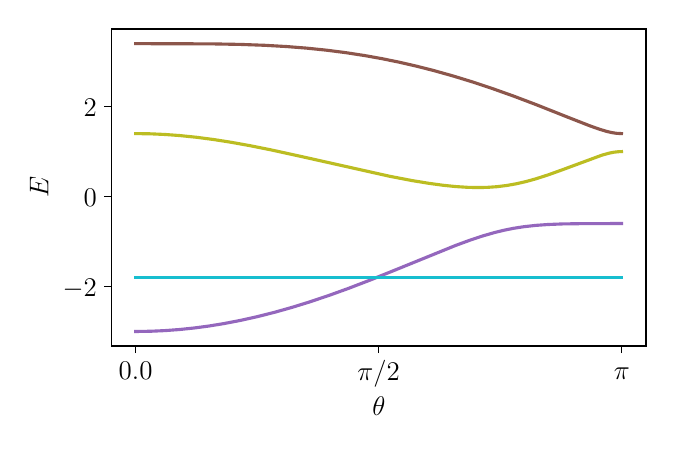}
           \includegraphics[width=0.3\textwidth]{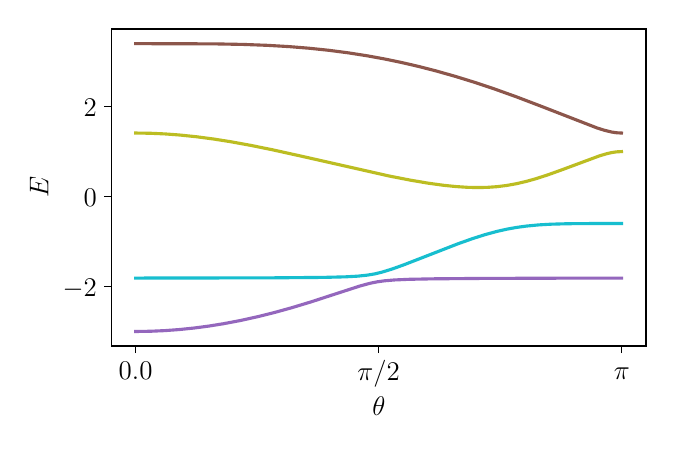}
      \includegraphics[width=0.3\textwidth]{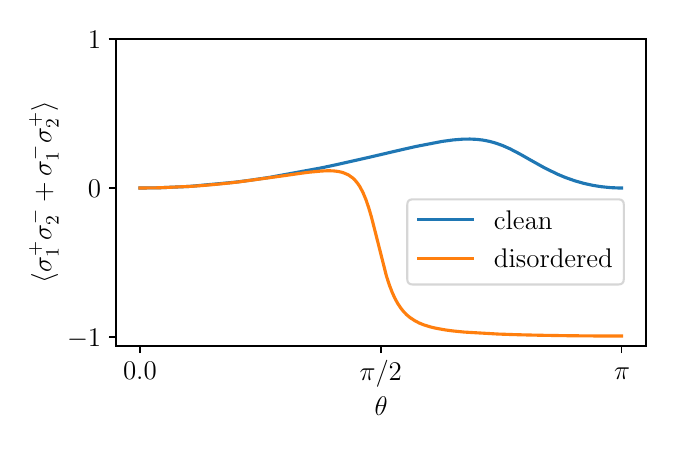}
  \caption{From [24]. Comparison of energy levels in the time-evolved state starting from the ground state at the north pole for (Left) the clean case and (Middle) with a mass disorder. On the (Right), we compare the expectation value $\langle{\sigma_1^+ \sigma_2^- + \sigma_1^- \sigma_2^+}\rangle$ for the time-evolved state starting from the ground state at the north pole for the clean and the disordered case. Used parameters are $d=1.0$, $r_z=0.2$, $r_{xy} = 0.8$, furthermore $M=0.6$, $(\delta M_1,\delta M_2) = (-0.1,0.1)$ for (Middle) and the disordered graph in (Right) and $v = 0.001$ for (Right), where $\theta = vt$.}
  \label{fig:majorana:ST_check}
\end{figure*}

In Fig.~\ref{fig:majorana:ST_check} we compare the energy levels for the clean (Left) and the disordered case (Middle) for a sweep along $\theta$ in time. The initially second-lowest band here corresponds to the singlet state with constant energy. The crossing between the two lowest lying bands for the clean case cannot be realized by a time-dependent driving. In the disordered case with a small mass imbalance shown in the Middle, a small gap between the two bands opens, leading to a transition and therefore to a final state close to the singlet state. It is clear that the ground state at the south pole becomes the singlet state in the limit $\Delta M=\delta M_1-\delta M_2 \to 0$ for $r_{xy}>0$. This state can be reached when the driving is adiabatic, i.e., if the driving velocity is much smaller than the magnitude of the gap opened. This is the case for Fig.~\ref{fig:majorana:ST_check} where $v/d = 0.5\cdot10^{-3}$ and the gap between the two lowest energy bands in the Middle panel was quantified numerically as $\Delta E/d \approx 93\cdot10^{-3}$.
In Fig.~\ref{fig:majorana:ST_check} on the Right we show the expectation value $\langle{\sigma_1^+ \sigma_2^- + \sigma_1^- \sigma_2^+}\rangle$ in time, in the disordered case, demonstrating that the state indeed approaches the singlet state for which this quantity goes to $-1$. From the ground-state wave function, we can evaluate for $r_{xy}>0$
\begin{equation}\label{eq:majorana:parity_second_order}
  \langle\sigma_1^+ \sigma_2^- + \sigma_1^- \sigma_2^+\rangle = - 1 + \left(\frac{\Delta M}{4 r_{xy}}\right)^2,
\end{equation}
which for the parameters of  Fig.~\ref{fig:majorana:ST_check} (Right) evaluates to $\sim -0.996$, while from ED it evaluates to $\sim -0.992$.

In the region with positive transverse coupling, the disorder therefore introduces a fractionalization of the disordered partial Chern marker. 

\end{appendix}

\bibliography{literature}

\end{document}